\documentclass{elsart}

\usepackage{epsfig}

\usepackage{amssymb}
\newcommand{\pom}{I\!\! P}
\begin{document}

\begin{frontmatter}



\title{CDF results on diffraction from Run I\\
and plans for Run II\thanksref{*}}
\thanks[*]{Presented at the \protect{\em IX\protect$^{th}$ `Blois Workshop' 
on Elastic and Diffractive Scattering\protect},
Pruhonice (Prague), Czech Republic, 9-15 June 2001. This article is an 
updated version of a paper presented by this author at the {\em DIS-2001
IX$^{th}$ International Workshop on Deep Inelastic Scattering}, Bologna, 
Italy, 27 April - 1 May 2001. 
}


\author{Konstantin Goulianos}

{\large (Representing the CDF Collaboration)}

\address{The Rockefeller University\\1230 York Avenue, New York, NY 10021, USA}

{\em Email address:} dino@physics.rockefeller.edu

\begin{abstract}
% Text of abstract
Results on soft and hard diffraction obtained by the CDF Collaboration 
in Run I of the Fermilab Tevatron $\bar pp$ collider are reviewed and 
compared with results from the DESY $ep$ collider HERA and with 
theoretical expectations. In addition, the CDF program for diffractive 
studies in Run II is briefly reviewed with emphasis on the relevant 
detector upgrades and physics goals. 
\end{abstract}

\begin{keyword}
diffraction \sep Pomeron
\PACS 13.85.Ni, 13.85.Qk, 13.87.Ce, 12.38.Qk,12.40.Ni
\end{keyword}
\end{frontmatter}

% main text
\section{Introduction}
\label{intro}

The signature of a diffractive event in $\bar pp$ collisions is a
leading proton or antiproton and/or a rapidity gap, defined as a region of 
pseudorapidity, $\eta\equiv -\ln\tan\frac{\theta}{2}$, devoid of particles 
(see Fig.~1). 

\begin{figure}[h,t]
\centerline{\hspace*{1cm}\psfig{figure=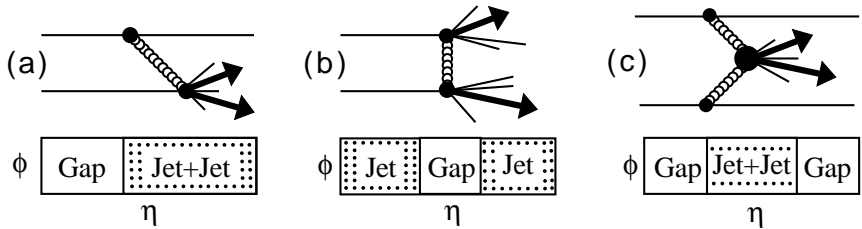,width=6in}}
\vglue -6.5in
\caption{Dijet production diagrams and event topologies for 
(a) single-diffraction,  (b) double-diffraction, and
(c)  double Pomeron exchange.}
\label{topology}
\end{figure}

In Run I of the Fermilab Tevatron $\bar pp$ collider, the 
CDF Collaboration studied the following 
diffractive processes:
\begin{itemize}
\item soft single-diffraction (SD) at $\sqrt s=$ 546 and 1800 GeV~\cite{CDF_SD} 
and soft double-diffraction (DD) at $\sqrt s=$ 630 
and 1800 GeV~\cite{CDF_DD}
\item $W$-boson~\cite{CDF_W}, dijet~\cite{CDF_JJG}, 
$b$-quark~\cite{CDF_b} and $J/\psi$~\cite{CDF_jpsi} production 
at $\sqrt s=$ 1800 GeV using rapidity gaps to identify diffractive events
\item dijets with a rapidity gap between jets at $\sqrt s=$ 
630~\cite{CDF_JGJ630} and 1800 GeV~\cite{CDF_JGJ1800_1,CDF_JGJ1800_2}
\item dijets with a leading antiproton 
at $\sqrt s=$ 630~\cite{CDF_JJ630} and 1800 GeV~\cite{CDF_JJ1800}
\item dijet production in double Pomeron exchange (DPE) with a leading 
antiproton and a rapidity gap on the proton side~\cite{CDF_DPE}
\end{itemize}
In addition to the normal variables used to discribe an interaction, two 
additional variables are needed to describe a diffractive event: the width of
the rapidity gap, $\Delta\eta$~\footnote{We use
pseudorapidity as an approximation to true rapidity.}, and 
the 4-momentum squared exchanged accross the gap, $t$. For single diffraction, 
$\Delta\eta\approx\ln\frac{1}{\xi}$, where $\xi$ is fractional momentum loss 
of the leading (anti)proton.

In Regge theory, which has traditionally been used to describe diffraction,
the rapidity gap is formed as a consequence of the nature of the 
exchanged Pomeron, which has the quantum numbers of the vacuum and therefore 
no hadrons are radiated in its exchange. Diffractive cross sections based on 
single Pomeron exchange factorize into two terms, one that has the form of 
a total cross section at the c.m.s. energy squared of the diffractive 
subsystem, $s'$, defined throught the equation $\ln s'=\ln s-\Delta\eta$, 
and a second term, 
which is a function of the diffractive variables $\Delta\eta$ and $t$.
The latter is usually referred to as `Pomeron flux' in single diffraction, or 
more generally as a `rapidity gap probability"~\cite{dino_flux,dino_gap}. 

In QCD, the {\em generic} Pomeron is 
a color-singlet combination of quarks and gluon with vacuum quantum numbers.
In addition to the question of Regge factorization, 
another question of interest for 
hard diffractive processes (those incorporating a hard 
scattering) is whether they obey QCD factorization. Comparisons 
among CDF results and between results from CDF and HERA show a rather severe 
breakdown of QCD factorization in diffraction. 

Below, we present the results obtained from the Run I studies 
and discuss briefly the CDF plans for diffractive physics in Run II.

\section{Soft diffraction}
Measurements of $pp$ and $\bar pp$ SD cross sections have shown that Regge 
theory correctly predicts the shape of the rapidity gap dependence for 
$\Delta\eta>3$, corresponding to a leading proton fractional momentum loss of 
$\xi\approx e^{-\Delta\eta}<0.05$, but fails to predict the 
correct energy dependence of the overall normalization, which 
at $\sqrt s=1800$ GeV is found to be suppressed by approximately an order of
magnitude relative to predictions based on 
factorization~\cite{CDF_SD,dino_flux,GM}. A new CDF measurement of the double 
diffraction differential cross section gives similar results (see Fig.~2).
\vglue -1em
\begin{center}
\begin{minipage}[t]{2.1in}
\phantom{x}
\vglue -1cm
\hspace*{-1cm}\psfig{figure=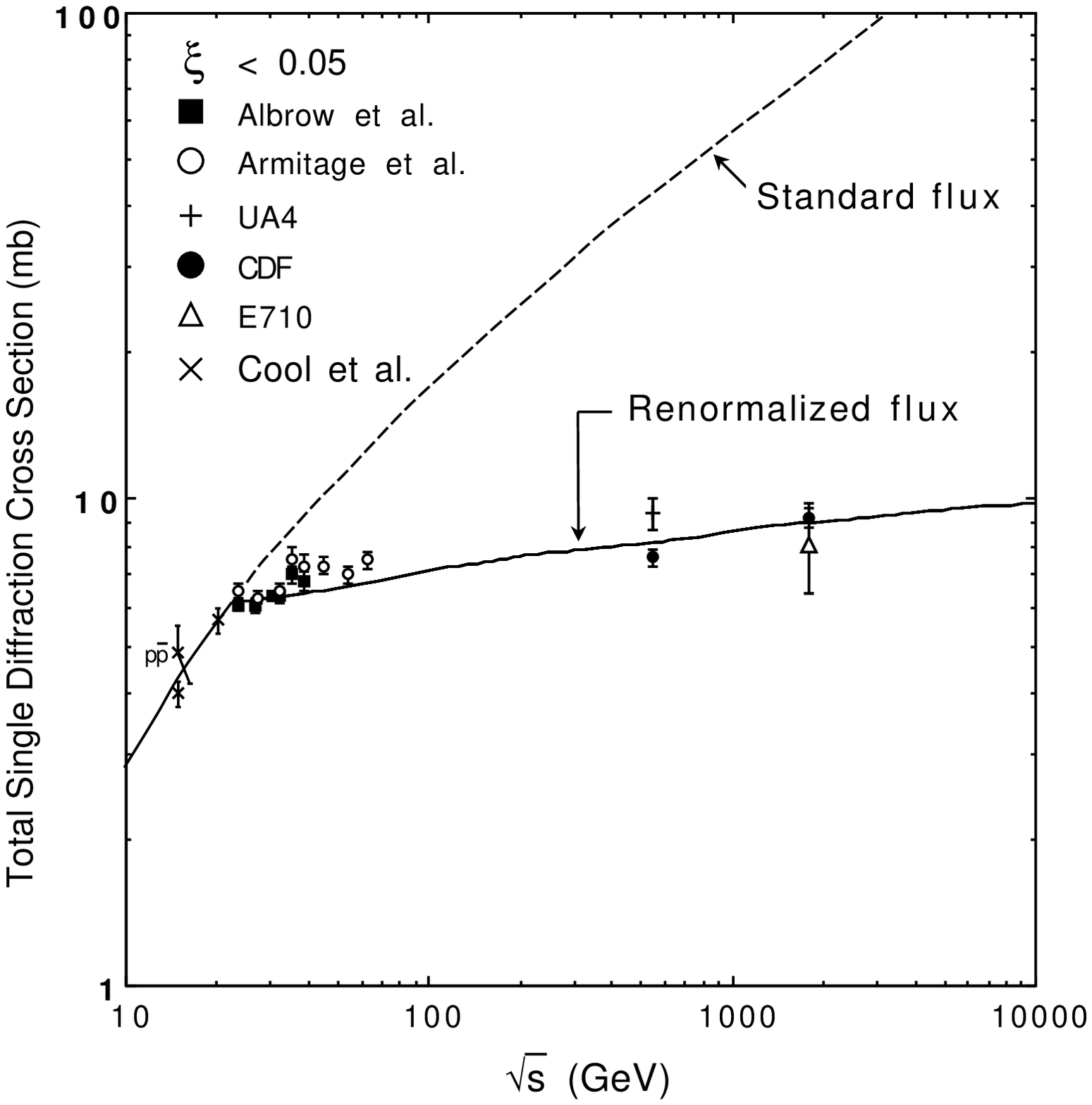,width=2.60in}
\vglue -2cm
{\small Fig. 2a: The $pp\,(\bar pp)$ $\sigma^t_{SD}$ vs $\sqrt s$.}
\end{minipage}
\hspace*{0.1cm}
\begin{minipage}[t]{2.25in}
\phantom{x}
\vglue -0.25cm
\psfig{figure=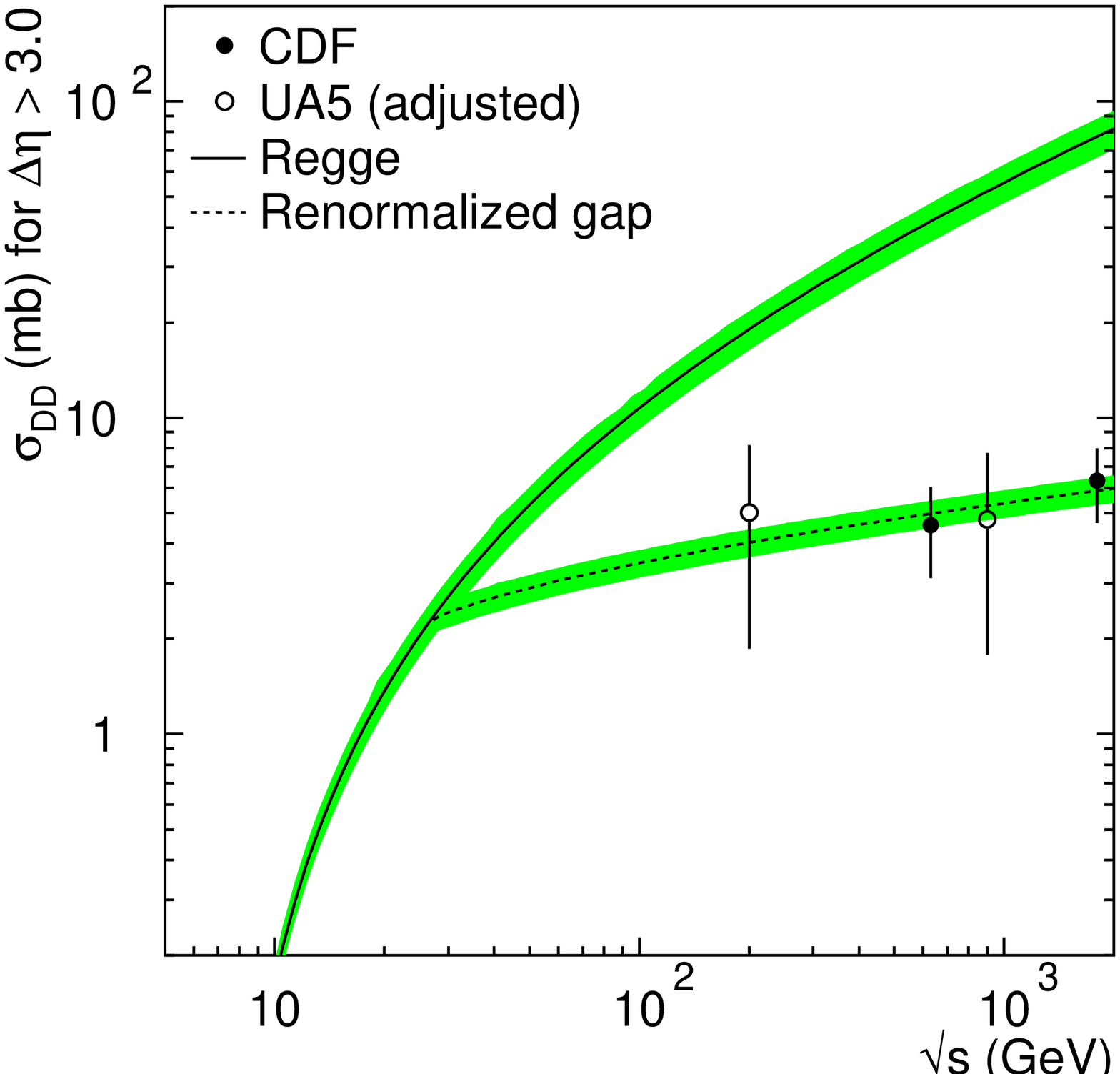,width=2.35in}
%\vglue 0.05cm
{\hspace*{0.8cm}{\small Fig. 2b: The $\bar pp$ $\sigma^t_{DD}$ 
vs $\sqrt s$.}}
%\vglue -1.9cm
%\hspace*{3cm}{\bf\large\it Preliminary}
\end{minipage}
\end{center}
\vglue 0.2cm

The SD and DD cross sections have very similar forms in terms of $\Delta\eta$:
\vspace*{-0.2cm}
\begin{eqnarray}
d^2\sigma_{SD}/dt\,d\Delta\eta=&
[Ke^{bt}e^{[2\alpha(t)-1]\Delta\eta}]\cdot 
[\kappa \beta^2(0)(s')^{\alpha(0)-1}]\\
d^3\sigma_{DD}/dt\,d\Delta\eta\,d\eta_c=&
[\kappa \;K\,e^{[2\alpha(t)-1]\Delta\eta}]
\cdot[\kappa \beta^2(0)(s')^{\alpha(0)-1}]
\end{eqnarray}
\vglue -0.2cm
Here, energy is measured in GeV,
$\alpha(t)=\alpha(0)+\alpha't$ is the Pomeron trajectory, 
$\beta(t)$ is the coupling of the Pomeron to the proton, $K=\beta^2(0)/16\pi$,
$\kappa=g_{\pom\pom\pom}/\beta(0)$, where $g_{\pom\pom\pom}$
is the triple-Pomeron coupling, $e^{bt}$ is the square of the proton 
form factor, $\eta_c$ the center of the rapidity gap, and  
$s'\equiv M_1^2M_2^2$, where $M$ is the diffractive mass, represents 
the (reduced) 
$s$-value of the diffractive sub-system, since $\ln s'=\ln s-\Delta\eta$ 
is the rapidity space where particle production occurs.
The second factor in the equations 
can be thought of as the sub-energy total cross section, which allows the 
first factor to be interpreted as a rapidity gap probability, $P_{gap}$.
For SD, it has been shown that renormalizing the Pomeron flux to 
unity~\cite{dino_flux}, 
which is equivalent to normalizing $P_{gap}$ over all phase space 
to unity, yields the correct energy dependence. The new CDF results
show that this also holds for DD, as predicted 
by a generalization of the Pomeron flux renormalization model~\cite{dino_gap}.
\section{Hard diffraction using rapidity gaps}
Using forward rapidy gaps to tag diffractive events, CDF 
measured the ratio of SD to non-diffractve (ND) rates for 
$W$-boson~\cite{CDF_W}, dijet~\cite{CDF_JJG}, 
$b$-quark~\cite{CDF_b} and $J/\psi$~\cite{CDF_jpsi} production 
at $\sqrt s=$ 1800 GeV, and using central gaps determined the fraction of 
jet-gap-jet events as a function of $E_T^{jet}$ and of rapidity gap separation 
between the two jets ($\Delta\eta^{jet}$) at $\sqrt s=$ 630 and 1800 GeV.

Forward gaps are defined as no hits in one of the beam-beam counters (BBC),
covering the region $3.2<|\eta|<5.9$, and no towers with energy
$E>1.5$ GeV in the forward calorimeter (FCAL, $2.4<|\eta|<4.2$) adjacent to 
the BBC with no hits. Using the POMPYT Monte Carlo simulation~\cite{POMPYT} 
with a flat gluon or quark Pomeron structure to generate diffractive events,
the measured SD/ND ratios were corrected for `gap acceptance', defined as 
the ratio of diffractive events with a gap to all diffractive events 
generated with $\xi=x_{\pom}<0.1$ in the selected kinematical range 
of the hard scattering products. 
For jet-gap-jet events, the gap was defined as no tracks or calorimeter 
towers with energy above $\sim 300$ MeV in the region $|\eta|<1$.
The ND background was estimated using events with both jets 
at positive or negative $\eta$.
\vglue 1em
\centerline{\small Table 1: Ratios of diffractive ($\xi<0.1$) to 
non-diffractive rates.}
\vglue -1in
\centerline{\psfig{figure=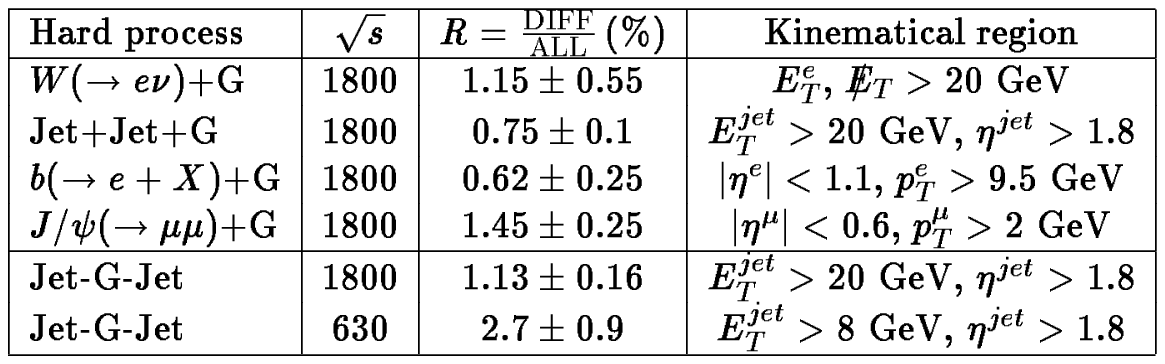}}
\vglue -8.45in

The results are summarized in Table~1. At $\sqrt s $=1800 GeV 
the DIFF/ALL ratios are of order 1\%. Since the processes under 
study have different sensitivities 
to the quark and gluon content of the Pomeron, the near equality of the 
SD to ND ratios indicates that the value of the gluon fraction in the Pomeron, 
$f_g^{\pom}$, 
is not very different from that in the proton. From the $W$, dijet and 
$b$-quark ratios, $f_g^{\pom}$ was determined to be 
$0.54^{+0.16}_{-0.14}$~\cite{CDF_b}. In addition, a suppression of a factor 
$D=0.19\pm 0.04$ was found in these ratios relative to POMPYT 
predictions using the standard Pomeron flux. Given that the POMPYT predictions
for dfiffractive processes at HERA are approximately correct, the 
observed large discrepancy between data and POMPYT predictions at the Tevatron 
indicates a breakdown of QCD factorization. 
The value of $D$ is approximately the same as that in soft SD (see Fig.~2), 
as was predicted in Ref.~\cite{dino_flux}.

An independent determination of $f_g^{\pom}$ was performed by comparing 
the measured $J/\psi$ DIFF/ALL ratio with that of dijet producion 
in association with a leading antiproton (discussed in the next section). 
This comparison, which was made at the same $x_{bj}$ (Bjorken $x$) value 
of the parton in the diffracted (surviving) nucleon, 
yielded $f_g^{\pom}=0.59\pm 0.15$ at $\langle x_{bj}\rangle=0.0063$, 
in agreement with the value obtained from the $W$, dijet and $b$-quark 
rapidity gap measurements. 

The double ratio of $J/\psi$ to $b$-quark DIFF/ALL ratios is $2.34\pm 0.35$.
Since both processes are mainly sensitive to the
gluon content of the Pomeron, CDF examined~\cite{CDF_jpsi}  
whether the difference 
in the two ratios could be attributed to the different 
average $x_{bj}$ values of the two measurements. 
Given the dependence $x_{bj}^{-0.45}$ of the measured
diffractive structure function~\cite{CDF_JJ1800} (see next section), 
the $J/\psi$ to $b$-quark double ratio is expected to be equal to 
$(x_{bj}^{J/\psi}/x_{bj}^{\bar bb})^{-0.45}$.
Since in these measurements only central $J/\psi$ or $b$-quark production
was considered, the ratio
$x_{bj}^{J/\psi}/x_{bj}^{\bar bb}$ is approximately proportional to the
ratio of the corresponding average $p_T$ value for each process, which is
$\approx 6$ GeV/$c$ for the $J/\psi$ and $\approx 36$ GeV/$c$ for
the $b$-quark (about three
times the average $p_T$ of the $b$-decay electron).
The expected value
for the double ratio is then $\approx (6/36)^{-0.45}=2.2$,
in agreement with the
measured value of $2.34\pm0.35$.

The ratio of jet-gap-jet fractions at $\sqrt s=630$ to 1800 GeV is 
$2.4\pm 0.8$. The $\Delta\eta^{jet}$, $E_T^{jet}$ and $x$-Bjorken
distributions are consistent with being flat~\cite{CDF_JGJ1800_2}. 
\section{Hard single diffraction using a leading antiproton spectrometer}
Using a Roman pot spectrometer to detect leading antiprotons 
and determine their momentum and polar angle (hence the $t$-value),
CDF measured the ratio of SD to ND dijet production rates 
at $\sqrt s$=630~\cite{CDF_JJ630} and 1800 GeV~\cite{CDF_JJ1800} as a 
function of $x$-Bjorken of the struck parton in the $\bar p$. In leading order
QCD, this ratio is equal to the ratio of the corresponding 
structure functions. For dijet production, the relevant structure function is
the color-weighted combination of gluon and quark terms given by 
$$F_{jj}(x)=x[g(x)+\frac49\sum_iq_i(x)]$$ 
The diffractive structure function,
$\tilde{F}_{jj}^D(\beta)$, where $\beta=x/\xi$ is the momentum fraction of the 
Pomeron's struck parton and the tilde over the $F$ indicates integration 
over $t$ and $\xi$, is obtained by multiplying 
the ratio of rates by the known $F_{jj}^{ND}$ and changing variables 
from $x$ to $\beta$ using $x\rightarrow \beta\xi$.

Results for $\sqrt s=$ 1800 GeV are presented in Fig.~3.
The comparison of $F^D_{jj}(\beta)$ with predictions based on 
diffractive parton densities extracted from DIS at HERA confirms 
the breakdown of factorization observed in the 
rapidity gap data presented in section 3. The difference in suppression 
factors between the rapidity gap and Roman pot data 
can be traced back to differences in kinematical acceptance.

\centerline{\psfig{figure=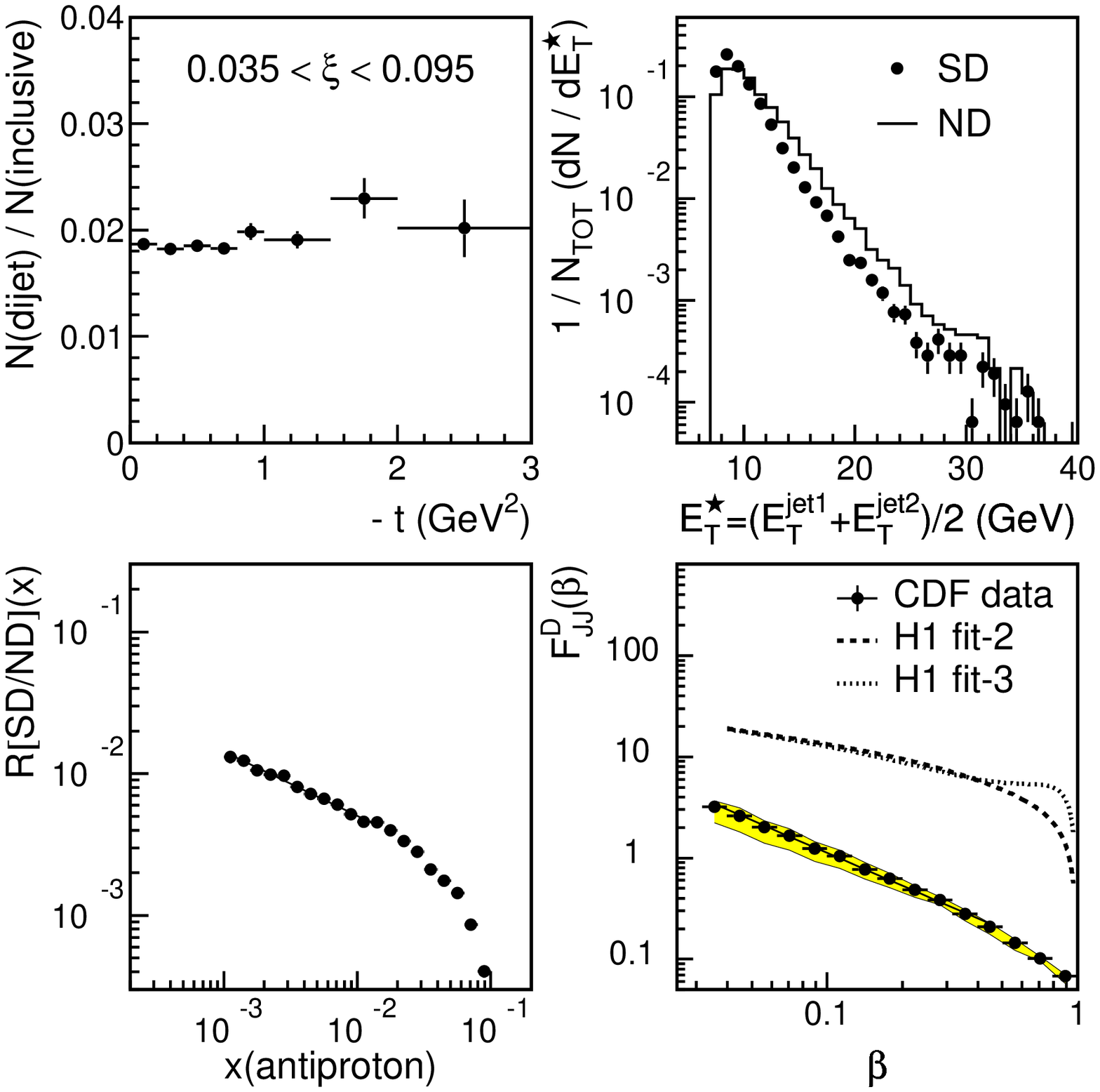,width=10cm}}
{\small Fig.~3: Inclusive and dijet 
diffractive results at $\sqrt s=$1800 GeV:\\
{\em (top left)} the ratio 
of dijet to inclusive SD event rates is independent of $t$;\\
{\em (top right)}
the $E_T^{jet}$ distribution is slightly steeper for SD than for ND events;\\
{\em (bottom left)} the ratio of SD to ND rates goes as
$x^{-0.45}_{bj}$ for $x<0.5\xi$;\\
{\em (bottom right)} the CDF diffractive structure function 
per unit $\xi$ is steeper than and severely suppressed relative to 
predictions based on diffractive parton densities 
extracted by the H1 Collaboration from DIS measurements at HERA.}
%\end{minipage}
\vglue 1em
To further characterize the diffractive structure function, CDF measured
its $\beta$ dependence as a function of $\xi$ and its $\xi$ dependence
as a function of $\beta$. In the region $\beta<0.5$
and $0.035<\xi<0.095$, the data are well represented
by the factorizable form
$$F_{jj}^D(\beta,\xi)=C\cdot \beta^{-n}\cdot \xi^{-m}$$
with $n=1.0\pm 0.1$ and $m=0.9\pm 0.1$, respectively, where the errors
are mainly due to the systematic uncertainty associated with the
measurement of $x$-Bjorken of the struck parton of the antiproton.
The observed $\xi$ dependence is much steeper than that
of the inclusive SD data sample, which in this $\xi$ region is 
approximately flat~\cite{GM}.
In Regge theory, the flat shape of the inclusive $dN/d\xi$
distribution in this region 
results from the superposition of a Pomeron exchange contribution,
which has a $\xi^{-\alpha(0)}\approx \xi^{-1.1}$ dependence, and a
Reggeon exchange contribution, which enters with an effective pion
trajectory~\cite{GM} and is $\sim \xi$. The measured value of
$m=0.9\pm 0.1$ indicates that dijet production is dominated by Pomeron exchange.
Such behaviour is expected in models in which the structure of the 
Pomeron is effectively built from the ND parton densities by
two exchanges, one at the high $Q^2$ scale
of the hard scattering and the other at the hadron mass scale of
order $1\;{\rm GeV}^2$~\cite{KKMR,softcolor,KG_JPG}. 

Diffractive dijet production was also studied at 
$\sqrt s$=630 GeV~\cite{CDF_JJ630}.
The diffractive structure function was extracted using the same method 
as at $\sqrt s$=1800 GeV, and the measurements of $F^D_{jj}(\beta,\xi)$ 
at the two c.m.s. energies were compared to test factorization. 
In the kinematical region of $E_T^{jet1,2}>7$ GeV, $E^*_T\equiv (E_T^{jet1}+
E_T^{jet2})/2>10$ GeV, $0.035<\xi<0.095$ and $|t|<0.2$ GeV$^2$,
the 630 to 1800 GeV ratio of the structure functions was found to be 
$$R=\frac{F^D_{jj}(\beta,\xi)|_{630\,{\rm GeV}}}
{F^D_{jj}(\beta,\xi)|_{1800\,{\rm GeV}}}=1.3\pm 0.2{\rm (stat)}^{+0.4}_{-0.3}
{\rm (syst)}$$
Within the quoted uncertainties, this ratio is compatible
with phenomenological predictions based
on the Pomeron flux renormalization~\cite{dino_flux}
and gap survival models~\cite{KKMR}.

\section{Dijet production in double-Pomeron exchange}
Factorization was also tested by comparing the 
ratio of DPE to SD dijet production rates to that of SD to ND.
Figure~4 illustrates the event topologies in pseudorapidity space 
and the corresponding Pomeron exchange diagrans for the two 
diffractive processes.
The comparison of the cross section ratios tests whether the 
normalization of the diffractive structure function of one of the nucleons 
is affected by the 
presence of a rapidity gap associated with the other nucleon. 

\centerline{\psfig{figure=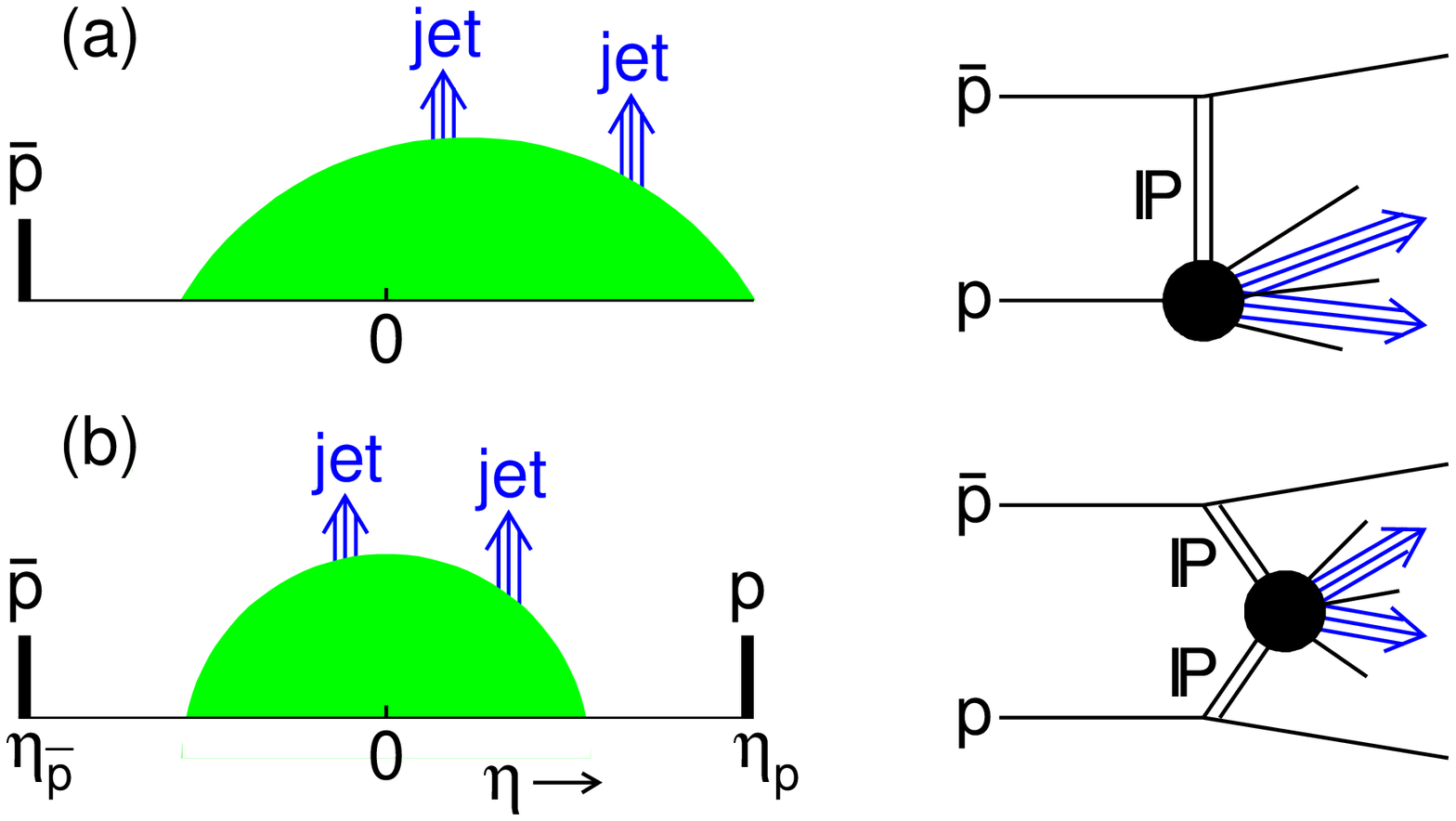,width=3in}}
\vglue -3cm
{\small Fig.~4: Illustration of event topologies in pseudorapidity, 
$\eta$,
and associated Pomeron exchange
diagrams for dijet production in (a) single diffraction and (b) double
Pomeron exchange. The shaded areas on the left side represent
particles not associated with the jets (underlying event).}
\vglue 1em
The DPE events were extracted from the leading antiproton 
data by requiring a forward rapidity gap on the 
proton side using the gap definition given in section 3. 
At $\langle \xi\rangle=0.02$ and 
$\langle x_{bj}\rangle =0.005$, the double ratio of SD/ND to DPE/SD rates, 
normalized per unit $\xi$, was found to be $0.19\pm 0.07$,
violating factorization~\cite{CDF_DPE}. 

	A search for exclusive dijet production in DPE, 
$p+\bar p\rightarrow p'+(jet1+jet2)+\bar p'$, yielded an upper limit 
of 3.7 nb for $0.035<\xi(\bar p)<0.095$ and jets of $E_T>7$ GeV and 
$\eta<1.7$~\cite{CDF_DPE}.

\section{Conclusions from Run I}
The CDF Run I diffractive studies revealed the following features 
regarding the process dependence of rapidity gap formation:\\
(1) In soft single and double diffraction, Regge factorization is violated 
in such a way as to lead to a scaling behaviour expressed as 
$s$-independence of the $M^2$ distribution of the differential cross sections.\\
(2) In hard diffraction, a severe breakdown of factorization is observed,
expressed as a supression of the the diffractive to non-diffractive 
production rates relative to predictions from Regge-type models based 
on factorization or from diffractive parton densities measured at HERA. 
The suppression factor is approximately equal to 
that observed in soft diffraction.\\
(3) The diffractive to non-diffractive productions rates are approximately 
flavour independent.\\
The above features lead to the conclusion that 
the probability of diffractive rapidity gap formation is, 
to first order,  process and flavor independent.

\section{Plans for Run II}
The CDF program for diffractive studies in Run II will include:\\
{\large (a) Hard single diffraction}\\
\hspace{1cm}-- Process dependence of $F^D$ (compare at the same $\xi$ and 
$x_{bj}$)\\ 
\hspace{1cm}-- $Q^2$ dependence of $F_{jj}^D$\\
{\large (b) Double Pomeron exchange}\\
\hspace{1cm}-- Soft DPE\\
\hspace{1cm}-- $F_{jj}^D(x_p)$ versus width of gap on the $\bar p$ side\\
\hspace{1cm}-- Exclusive dijet and $\bar bb$ production\\
\hspace{1cm}-- Low mass exclusive states (glueballs?)\\
{\large (c) Hard double diffraction}\\
\hspace{1cm}-- jet-gap-jet events at high $\Delta\eta^{jet}$ (test BFKL)\\
{\large (d) Unexpected discoveries!}
\vglue 0.1in
The Run II program will be implemented by upgrading CDF to include the forward 
detector system shown schematically in Fig.~4. This system comprises:
1. A Roman Pot Specrometer (RPS) on the antiproton side
to detect leading antiprotons and measure $\xi$ and $t$\\
2. {Beam Shower Counters (BSC) covering the region $5.5<|\eta|<7.5$
to be used for triggering on events with forward rapidity gaps}\\
3. {Two `MiniPlug' calorimeters in the region $3.5<|\eta<5.5$}\\
\vglue -0.35in
\centerline{\psfig{figure=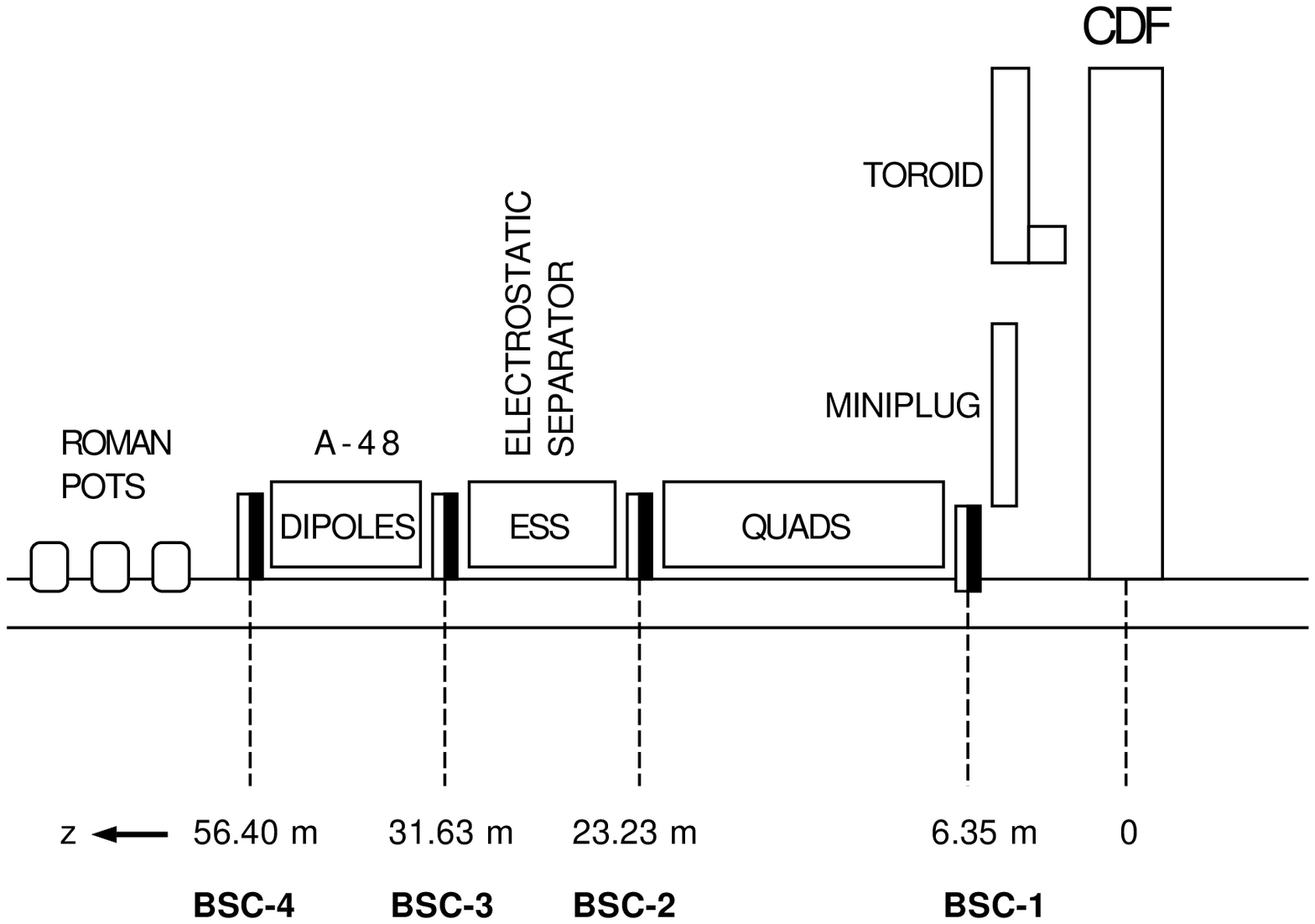,width=4in}}
\vglue -2in
\centerline{\small Fig. 5: CDF forward detectors for Run II}
\vglue 0.1cm
The RPS and BSC systems are already installed, and the MiniPlug 
installation is scheduled for October 2001.


\begin{thebibliography}{00}

\bibitem{CDF_SD}F. Abe {\em et al.}, Phys. Rev. D {\bf 50}, 5535 (1994).
%
\bibitem{CDF_DD}{\em Double diffraction dissociation at the Fermilab Tevatron 
collider}, CDF Collaboration, accepted by Phys. Rev. Letters.
%
\bibitem{CDF_W}F. Abe {\em et al.}, Phys. Rev. Lett. {\bf 78}, 2698 (1997).
%
\bibitem{CDF_JJG}F. Abe {\em et al.}, Phys. Rev. Lett. {\bf 79}, 2636 (1997).
%
\bibitem{CDF_b}T.~Affolder {\em et al.}, Phys. Rev. Lett. {\bf 84}, 232 (2000).
%
\bibitem{CDF_jpsi}{\em Observation of diffractive $J/\psi$ production at 
the Fermilab Tevatron}, CDF Collaboration, submitted to
Phys. Rev. Letters.
%
\bibitem{CDF_JGJ630} F. Abe {\em et al.},
Phys. Rev. Lett. {\bf 80}, 1156 (1998); {\bf 81}, 5278 (1998).
%
\bibitem{CDF_JGJ1800_1}F. Abe {\em et al.}, 
Phys. Rev. Lett. {\bf 74}, 855 (1995).
%
\bibitem{CDF_JGJ1800_2} F. Abe {\em et al.}, 
Phys. Rev. Lett. {\bf 80}, 1156 (1998).
%
\bibitem{CDF_JJ630}{\em Diffractive dijet production at $\sqrt s=630$ and 1800 
GeV at the Fermilab Tevatron}, CDF Collaboration, to be submitted to 
Phys. Rev. Letters.
%
\bibitem{CDF_JJ1800}T.~Affolder {\em et al.},
Phys. Rev. Lett. {\bf 84}, 5043 (2000).
%
\bibitem{CDF_DPE} T.~Affolder {\em et al.},
Phys. Rev. Lett. {\bf 85}, 4215 (2000).
%
\bibitem{dino_flux}K. Goulianos, Phys. Lett. B 358, 379 (1995).
%
\bibitem{dino_gap}K. Goulianos, ``Diffraction: Results and Conclusions",
in {\em Proceedings of
Lafex International School of High Energy Physics, Rio de Janeiro, Brazil,
February 16-20 1998}, edited by
Andrew Brandt, H\'{e}lio da Motta and Alberto Santoro; hep-ph/9806384.
%
\bibitem{GM}K. Goulianos and J. Montanha,
Phys. Rev. D {\bf 59}, 114017 (1999).
%
\bibitem{POMPYT}  P. Bruni, A. Edin and G. Ingelman,
Report No. DESY-95, DRAFT, ISSN 0418-9833; http://www3.tsl.uu.se/thep/pompyt/
%
\bibitem{KKMR}A.B. Kaidalov, V.A. Khoze, A.D. Martin 
and M.G. Ryskin, arXiv:hep-ph/0105145 
(and private communication with the authors).
%
\bibitem{softcolor}R. Enberg, G. Ingelman and N. Timneanu,
J. Phys. G {\bf 26}, 712 (2000) [arXiv:hep-ph/0001016]; see also
arXiv:hep-ph/0106247.
%
\bibitem{KG_JPG}K. Goulianos,
J. Phys. G: Nucl. Part. Phys. {\bf 26}, 717 (2000).
%
% notes:

% \bibitem{label2}
% If there is a note, it should come last:
% \bibitem{label3} \note
% \end{subbibitems}

%\bibitem{}

\end{thebibliography}
\end{document}